\documentclass[12pt,preprint]{aastex}

\newcommand{\ms}{\mbox{m s$^{-1}~$}}
\newcommand{\mse}{\mbox{m s$^{-1}$}}

\newcommand{\msun}{M$_{\odot}$}

\newcommand{\mjup}{M$_{\rm JUP}$}

\newcommand{\mearth}{M$_{\earth}~$}

\newcommand{\msini}{$M \sin i~$}

\lefthead{Arriagada {\it et~al.\/}}
\righthead{New Magellan Planets}
\received{}
\accepted{}

\begin{document}

\title{Two planetary companions around the K7 dwarf GJ 221 :  a hot
super-Earth and a candidate in the sub-Saturn desert range}

\author{Pamela Arriagada\altaffilmark{2,3},
Guillem Anglada-Escud\'e\altaffilmark{3,6}, 
R. Paul Butler\altaffilmark{3}, 
Jeffrey D. Crane\altaffilmark{4}, 
Stephen A. Shectman\altaffilmark{4}, 
Ian Thompson\altaffilmark{4}, 
Sebastian Wende\altaffilmark{6}, 
Dante Minniti\altaffilmark{2,5}
}

\authoremail{parriaga@astro.puc.cl}

\altaffiltext{1}{Based on observations obtained with
the Magellan Telescopes, operated by the Carnegie
Institution, Harvard University, University of Michigan,
University of Arizona, and the Massachusetts Institute
of Technology.} 

\altaffiltext{2}{Department of Astronomy, Pontificia
Universidad Cat\'olica de Chile, Casilla 306, Santiago 22, Chile}

\altaffiltext{3}{Department of Terrestrial Magnetism, Carnegie
Institution of Washington, 5241 Broad Branch Road NW,
Washington D.C. USA 20015-1305}

\altaffiltext{4}{The Observatories of the Carnegie Institution of Washington,
813 Santa Barbara Street, Pasadena, CA USA 91101}

\altaffiltext{5}{Vatican Observatory, V00120 Vatican City State, Italy}

\altaffiltext{6}{       Universit\"{a}t G\"{o}ttingen,
       Institut f\"ur Astrophysik,
       Friedrich-Hund-Platz 1,
       37077 G\"{o}ttingen, Germany
}

\begin{abstract}

We report two low mass companions orbiting the nearby K7 dwarf GJ 221 that have
emerged from re-analyzing 4.4 years of publicly available HARPS spectra
complemented with 2 years of high precision Doppler measurements with
Magellan/PFS. The HARPS measurements alone contain the clear signal of a low
mass companion with a period of 125 days and a minimum mass of 53.2 \mearth (GJ
221b), falling in a mass range where very few planet candidates have been found
(sub-Saturn desert). The addition of 17 PFS observations allow the confident
detection of a second low mass companion (6.5 \mearth) in a hot orbit (3.87 days
period, GJ 221c). Spectrocopic and photometric calibrations suggest that GJ 221
is slightly depleted ([Fe/H]$\sim$ -0.1) compared to the Sun so the presence of
two low mass companions in the system confirms the trend that slightly reduced
stellar metallicity does not prevent the formation of planets in the super-Earth
to sub-Saturn mass regime.

\end{abstract}


\section{Introduction}
\label{intro}

After more than a decade of planet search discoveries, the current planetary
candidate census has extended to more than 700 planets and counting. These have
been discovered using a range of techniques including radial velocity
\citep{mayor:1995,butler:2006,udry:2007}, photometry of transiting planets
\citep{henry:2000,char:2000}, microlensing \citep{bennet:2009,gaudi:2010},
direct imaging \citep{kalas:2008,marois:2008,lafre:2011} and transit
timing variatons \citep{nesvorny:2012}. Up to date no planets have been found using astrometry, although the astrometric signal of already known planets has been detected \citep{benedict:2002,mcarthur:2004,bean:2007,reffert:2011}. 

The Kepler mission has recently identified more than two thousand planet
candidates \citep{bathala:2013}, including several in the terrestrial mass range. Current and future transiting surveys that have been designed to find habitable planets around nearby stars are the ground-based MEarth project, and the Transiting Exoplanet Survey Satellite. Still, ground-based precision RV surveys remain the only technique that has been able to
find terrestrial mass planets around nearby stars in their habitable zones \citep{udry:2007, vogt:2010, anglada:2012b}.  Recently, much of the RV
effort has been directed toward the detection of low-mass planets around the lowest-mass stars: 
for a given planetary mass and period, a low-mass M-dwarf host will show a larger
Doppler amplitude than a G-dwarf host, making lower mass planets easier to detect.

Currently, the most precise Doppler planet search facilities can achieve an
internal long-term precision of 1--3 \ms \citep{mayor:2011,vogt:2010},
which is enough to detect short-period Earth-mass planets around M dwarfs. At
this stage, characterization of the intrinsic-stellar noise sources such as
jitter, convective granulation and asteroseismological p-mode oscillations are
becoming more important \citep{dumusque:2012, tuomi:2013}.

Consequently, the PFS Magellan Planet Search is targeting a sample of 500 of the
nearest stars ($< 100$ pc), 200 of which are chromospherically quiet late K and
early M dwarfs. The PFS Magellan Planet Search makes use of the Carnegie Planet
Finder Spectrograph (PFS). PFS was commissioned at the 6.5 meter Magellan
Clay telescope at Las Campanas Observatory (LCO) in Chile beginning in September 2009 and went into full operation in January 2010.
PFS  is a temperature controlled high resolution spectrograph \citep{crane:2006, crane:2008, crane:2010}. 
The spectrograph is maintained at a constant temperature ($\pm0.005^{\circ}$C) near 25$^{\circ}$C throughout the year so that the internal optical focus will remain constant without need for adjustment and the refractive index of the air will be
stable. An Iodine absorption cell \citep{marcy:1992} is mounted in front of
the instrument's entrance slit, imprinting the reference Iodine spectrum
directly on the incident starlight, providing both a wavelength scale and a measure of
the spectrometer point-spread-function \citep{butler:1996}. The Iodine cell is
a temperature controlled sealed pyrex tube, such that the column density of
Iodine remains constant indefinitely. PFS was built exclusively to obtain precision Doppler velocity
measurements of GKM stars.

With the unexpected diversity of extrasolar planets found to date, every new planet continues to contribute
to our understanding of planet formation.  In this sense, multiplanet systems are especially important, as they offer the best way to constrain evolutionary history by detailed orbital and dynamical analysis. In this work we present the detection of two low-mass companions to the nearby low-mass star GJ 221, by reanalyzing 61 HARPS high precision radial velocity measurements and including 17 PFS high precision radial velocity measurements. In Section 2 we describe the observations and data from both spectrometers. Section 3 describes the stellar properties of GJ 221. The orbital analysis is described in Section 4. Finally a discussion and conclusions are presented in Section 5.

\section{Observations}
\label{obs}

\subsection{PFS Observations and data}

Using a 0.5 arc-sec slit, PFS obtains spectra with a resolution of R $\sim$
80000 in the iodine region (5000-6300 \AA) and covers a wavelength range of 3880
to 6680 \AA. Only the iodine region is used in the Doppler analysis, and
the Ca II H and K lines are used to monitor stellar activity.

Total exposure times range from 300 seconds on the brightest objects to 720 seconds
on fainter ones. This scheme ensures adequate S/N in the iodine region
(S/N$>300$ per spectral resolution element) and guarantees proper cosmic
ray removal for the faintest targets since longer exposure times yield too many cosmic rays to be handled properly in an automated way. In the case of bright targets a minimum exposure time of 300 seconds allows proper time-averaging over unresolved low-order stellar p-modes.

Calibrations were acquired at the beginning and at the end of each night. These
include 30 flat field images with the slit illuminated by an incandescent lamp, two incandescent exposures passing through the iodine cell, using both the 0.3" and  0.5" slits, and two exposures of a rapidly rotating B star taken through the iodine cell.% near the meridian. 

%We also have ThAr shots which are currently not being used.

Extraction of the spectra from raw CCD images
was carried out using an IDL-based reduction pipeline that performs
flat-fielding, removes cosmic rays and measures scattered light from inter order
pixels and subtracts it.  %Cosmic rays are removed from 2D echellogram by estimating the intensity distribution of each pixel and comparing it with information from neighboring pixels. %check this with paul

The extracted spectra covers the full wavelength range of the spectrograph. No sky subtraction is done as the sky brightness around
the iodine region is negligible compared to the brightness of the sources.

Doppler shifts from the spectra are determined with the spectral synthesis
technique discussed in greater detail in \citet{butler:1996}. In the case of PFS,
the iodine region is divided into $\sim 800$ chunks of 2 \AA~each. Every
chunk produces an independent measure of the wavelength, PSF, and Doppler shift.
The final measured velocity is the weighted mean of the velocities of the
individual chunks. The internal uncertainties presented for all of the PFS RVs are derived as the uncertainty in the mean of the $\sim$800 velocities from one echelle spectrum. The derived RVs using PFS spectra are presented in table \ref{tab:PFSvels}.

\begin{figure}
\plotone{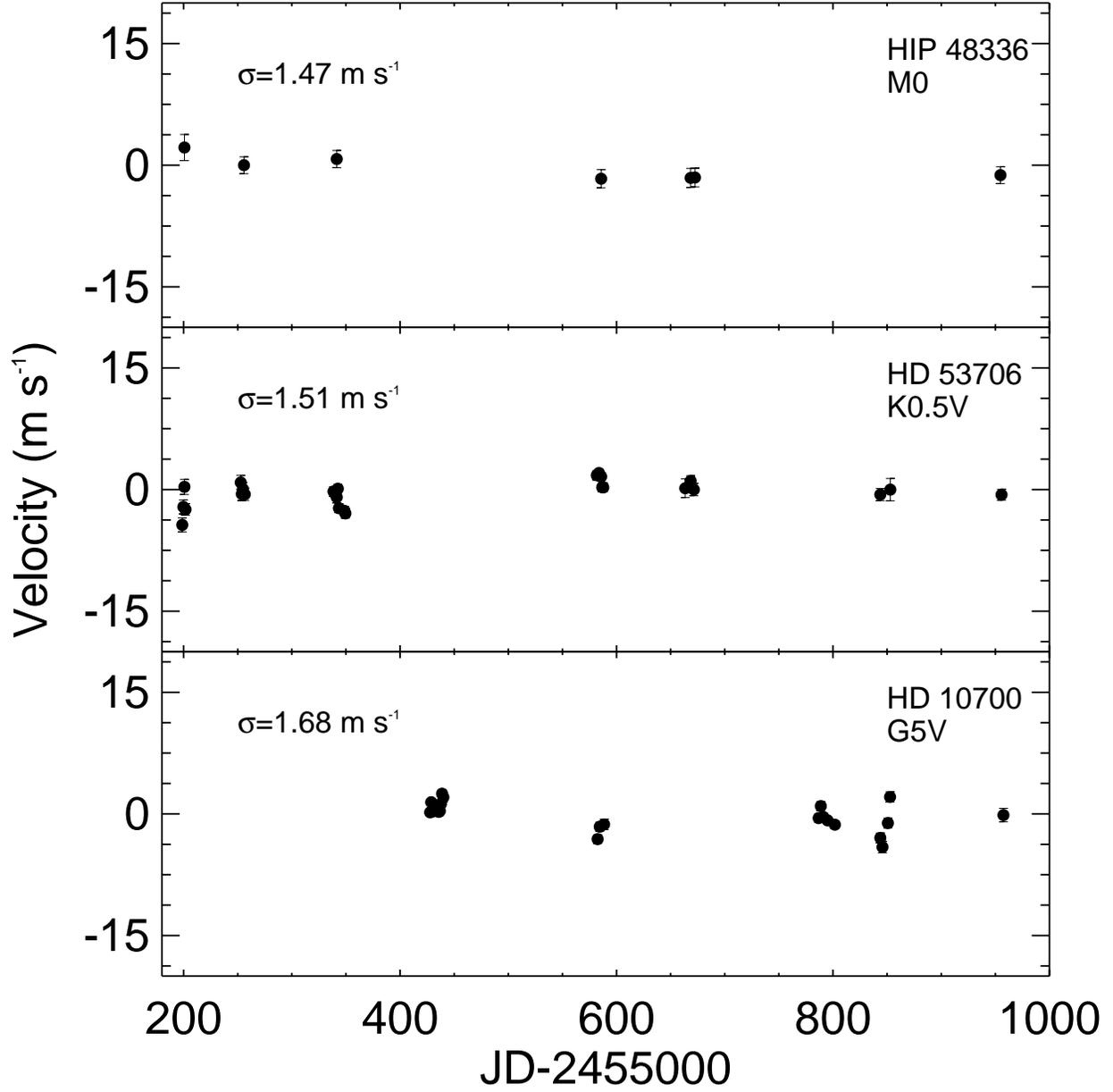}
\caption{Four PFS stable stars with spectra types ranging from mid G to early M.}
\label{fig:stable}
\end{figure}

Since the start of the project, we have monitored a number of stable main
sequence stars with spectral types from late F to early M. Figure \ref{fig:stable}
shows  the RVs of 3 of these stars during a period spanning more than 2 years of
observations. The Magellan/PFS system achieves measurement precision of 1.5
\mse, as demonstrated by this figure. The internal median uncertainty of the
observations corresponds to the best estimate for the uncertainty due to photon
statistics. For the stable stars (RMS $\sim$1.5 \ms and internal uncertainty of
$\sim$1.1 \mse), the total uncertainty due to systematic errors, stellar jitter,
and unknown planets is therefore $\sim$1 \ms ($\sqrt{1.5^2 - 1.1^2}$). In order
to mitigate at least one of these sources of error, stellar jitter, we have
selected the most chromospherically quiet stars and have chosen the observing
strategy described above.

%The internal uncertainties do not account for the other two main sources of error: photon noise and stellar jitter. %check this last sentence?!?

\subsection{HARPS public spectra}

As of April 2012, we found 61 public spectra of GJ221 in the HARPS-ESO database.
GJ 221 has been observed as part of various programs with integration times
between 300 and 900 seconds. RVs given by the archive which are obtained through
the HARPS pipeline contain several outliers. Apart from one
observation that clearly failed to converge (RV offset of -1500 km$^{-1}$) all 7
of the measurements taken within the program 074.C-0037 (Planets around young
stars) were processed with a G2 binary mask and show RV offsets of 500 \ms with
respect to the other ones. All the available spectra were consistently reanalyzed
using the HARPS-TERRA software \citep{anglada:2012a} using the standard setup
for M dwarfs (cubic blaze function correction, only echelle orders redder than
$\lambda>4400$ \AA~ were used which corresponds to echelle apertures from 22 to
71). The method consists of measuring relative RV offsets compared to a high
signal-to-noise ratio template built by co-adding all the available
observations. While the typical internal errors are of the order 1.2 m s$^{-1}$,
the derived RVs show a root-mean-squared (RMS) of 6.9 m s$^{-1}$, indicating a
significant excess of variability. HARPS-TERRA derived RVs are presented in 
table \ref{tab:TERRAvels}.

\section{Properties of GJ 221}
\label{stellarprop}

GJ 221 (HIP 27803) has been classified as both a K7 dwarf \citep{reid:1995} and
M0V \citep{upgren:1972} with an apparent visual magnitude of $V=9.69$ and a
color $B-V=1.35$.  The HIPPARCOS parallax \citep{hipparcos:2007} gives a
distance of 20.31 pc, yielding an absolute magnitude of $M_V=8.15$. This makes
the star slightly below the zero-age main sequence relation given in
\citet{wright:2004}, which would be compatible with a slightly metal-poor main
sequence star (metallicity between $-0.5$ and the solar one). According to the
UVW Galactic velocities \citep{hawley:1996}, GJ221 should belong to the young
disk, giving a lower threshold for its age of 0.1 Gyr. Such lower limit 
for the age matches the lower limit obtained from the (V-Ic)-age relations 
given by \citet{gizis:2002}, but it is not a very informative constraint. The
chromospheric activity indicator log(R'$_{\rm HK}$)=-4.8 suggests an age closer
to 4 Gyr \citep{mamajek:2008} and the lack of strong flaring events during 70+
observations also suggest an age older than 1 Gyr. As we discuss later, the
spectroscopic analysis is in agreement with evolutionary models if an age higher
than 1 Gyr is assumed. At least one direct metallicity measurement has been
reported before\citep{casagrande:2008}. It reported a metallicity of [Fe/H]=
-0.26. The internal scatter on their sample was rather large ($\sim$ 0.2 dex) so
such measurement must be taken with care.

\begin{figure}
%\plotone{stable1.ps}
\includegraphics[width=5.0in]{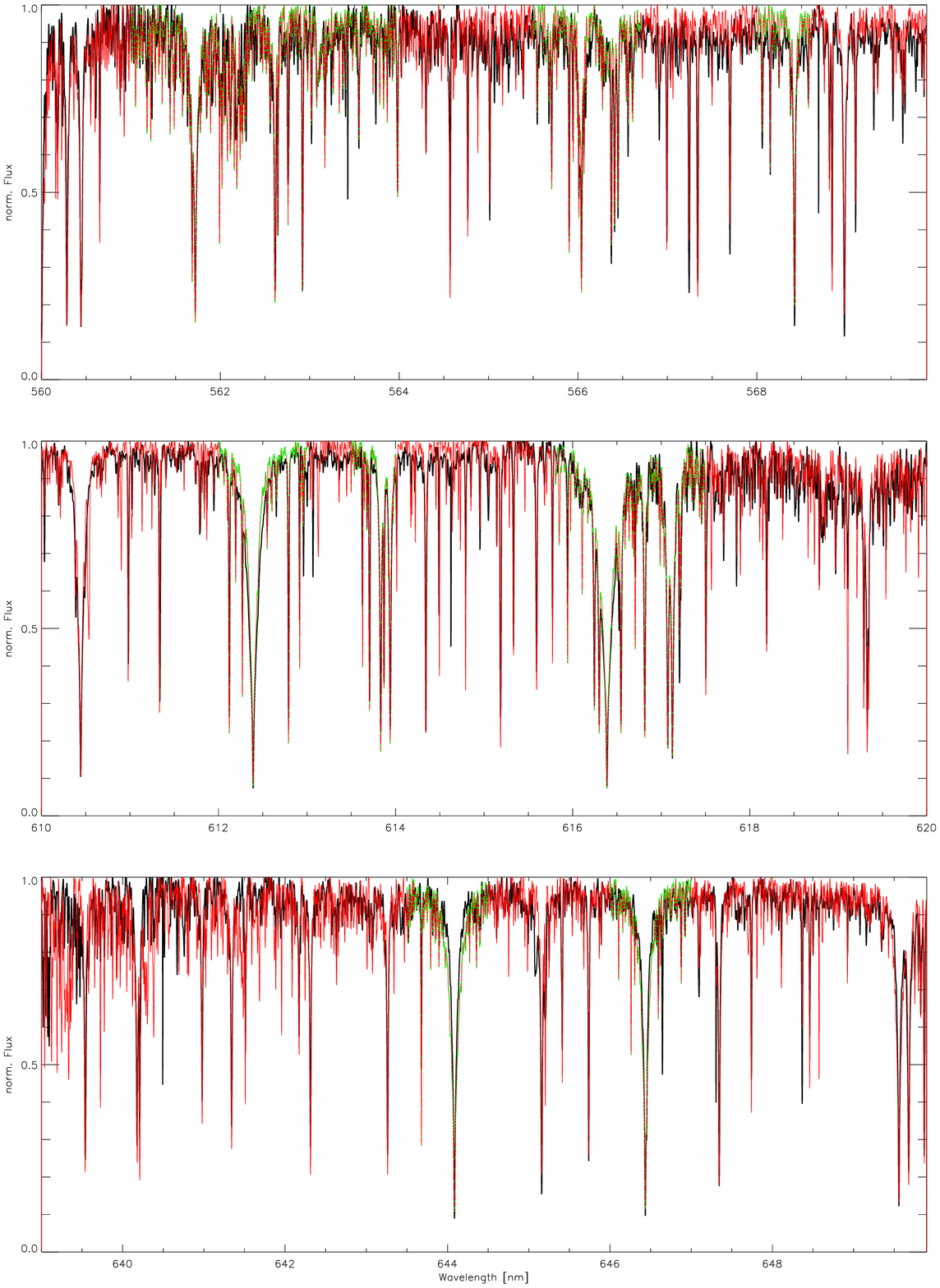}
\caption{Spectral lines used to derive the parameters of the star. The
wide/strong lines between 562nm and 568nm are FeI lines (top panel) and the wide
strong ones between 610 and 650nm are from CaI (second and third panels). Only
the regions marked in green where used to obtain the spectroscopic fit.}
\label{fig:spectrum}
\end{figure}

We first estimated the mass of GJ 221 using the mass-luminosity relation by
\citet{delfosse:2000} and the V and K photometric band values given in Table
\ref{tab:starparam} and obtained M$_*$ = 0.70 \msun. Assuming this mass, an age of
4 Gyr and the aforementioned metallicity, we use the evolutionary models in
\citet{baraffe:1998} to derive an effective temperature of $\sim$ 4300 K. This
value is too high compared to previous spectral type determinations (K7-M0 star
should correspond to T$_{\rm eff}\sim 4000$ K or lower). In order double check
the stellar parameters with an independent approach, we also derived some of
them spectroscopically. The fitting procedure consist on forward modeling
synthetic PHOENIX spectra to match the observations using a least-squares
solver. The new PHOENIX spectral library has been recently released and is
described in full detail on \citet{husser:2013}. In addition to flux
normalization for the continum and resolution of the instrument (instrumental
line-profile assumed Gaussian), the synthetic spectra also depends on the
parameters of interest : T$_{\rm eff}$, [Fe/H], log g and $v_{rot}\sin i$. The
analyzed spectrum was generated by coadding all the HARPS observations and has
an aproximate S/N of 400/pix at 6100 \AA. Because atmospheric models of cool
stars are still not very accurate in predicting spectra with regions heavy
blancketed by mollecular features, we focussed on small sub-sections of it
centered on a well known Fe and CaI lines (see Fig.~\ref{fig:spectrum}),
that are sensitive to $\log$ g and metallicity. The solution was
initialized close to our preliminary estimates (discussed before) and the
initial value of $\log g$ was taken from \citep[][ - who assumed solar
metallicity]{takagi:2011}. The solution converged to $\log g=4.5 \pm 0.1$,
T$_{\rm eff}=4040 \pm 50$ K and a metallicity of  [Fe/H] $= -0.08 \pm 0.1$. Also
a rotational velocity of $v_{rot}\sin{i}=1.8\pm0.1$\,km\,s$^{-1}$ was obtained
by fitting the spectra. Using these values, the evolutionary models yield a mass
of 0.63M$\sun$ if an age of $\sim 5$ Gyr was assumed (any age between 2 and 10
Gyr would fit equally well). This value is $\sim$ 11\% lower than the value
obtained from the mass-luminosity calibration which is known to be uncertain at
the 10\% level in this mass range \citep{delfosse:2000}. The star is right in
the middle of the transition form K to M spectral types and a small change in
the parameters leads to a sensible change in derived properties such as
luminosities and masses. All things considered, we think that the most
consistent picture is recovered when stellar mass of GJ 221 is obtained from the
spectroscopic adjustment + evolutionary models (0.65 \msun) and that 5$\%$
(0.032 \msun) is a realistic estimate of its uncertainty. The same interpolation
of the evolutionary models produces a luminosity of 0.095 L$_\odot$. Updated
star parameters and uncertainties are summarized in Table \ref{tab:starparam}.

\section{Orbital Analysis}

\begin{figure}
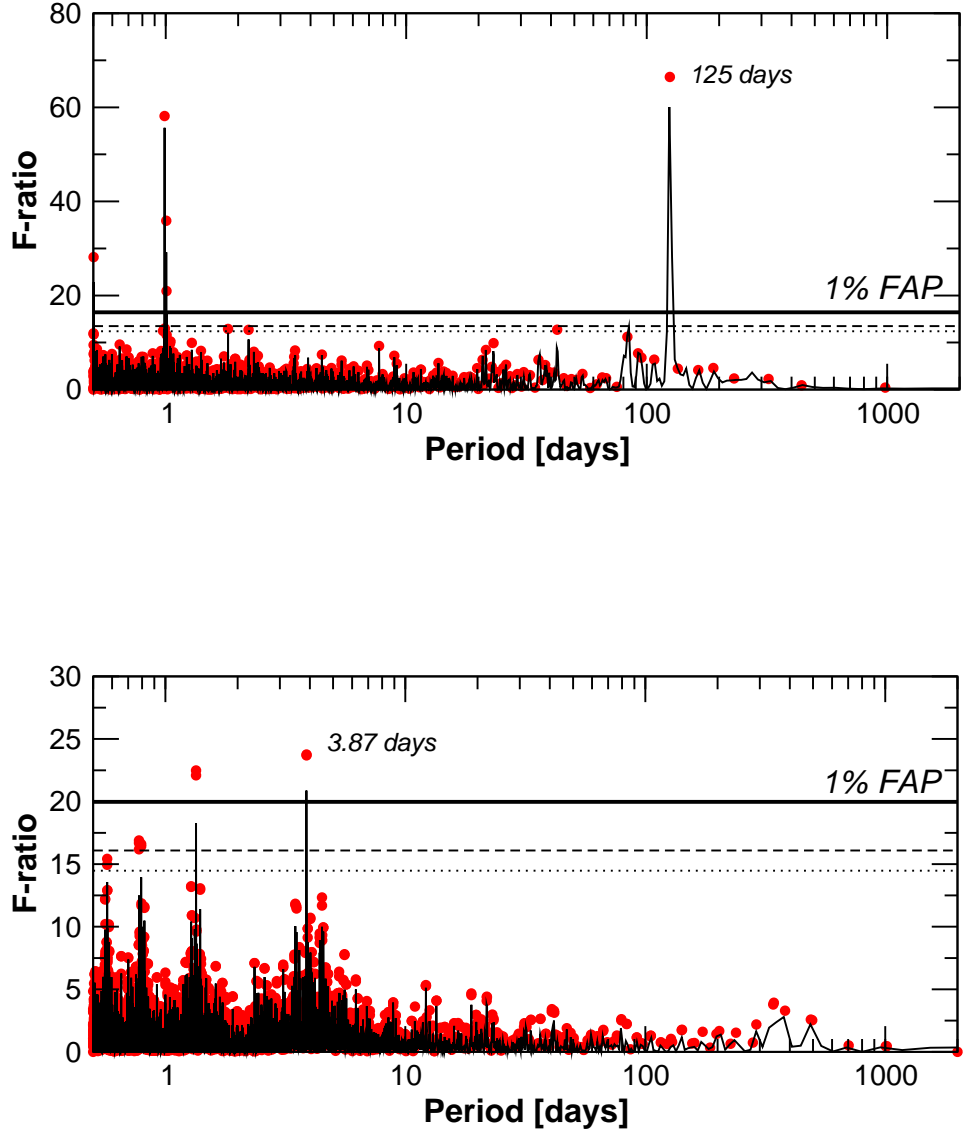

\includegraphics[width=5.0in]{GJ221_b_periodogram.eps}
\vskip1.0in
\includegraphics[width=5.0in]{GJ221_c_periodogram.eps}

\caption{Detection periodograms of the two candidate planets detected in the RV
measurements of GJ 221 (black lines). Red dots correspond to the F-ratios of the refined solutions. The signals are listed from top to bottom in order of detection. Strong aliases of both signals are found at the expected frequency shifts.}

\label{fig:periodograms}
\end{figure}

The final orbits were derived with our custom-made software (see description of
the methods below) and the solutions were double-checked with the SYSTEMIC package
\citep{meschiari:2009}. The detection false alarm probability
(FAP) of each candidate was evaluated as follows. First, a least-squares
periodogram \citep{cumming:2004,anglada:2012b} was computed on the residuals to
the k-planet fit. It is known that when using periodogram of the residuals,
correlations and aliases \citep[e.g.,][]{hd10180} can decrease the apparent
significance of additional low-amplitude signals in multi-planetary systems. To
account for this, the solution provided by the periodogram of the residuals is
refined at each test period by allowing the Keplerian parameters of all of the
already detected k-planets to adjust. The orbit for the (k+1) planet candidate
is kept circular but its orbital period, phase and amplitude are also allowed to
adjust ensuring that the optimal multi-planet solution is explored at each test
period. The powers (F-ratio) of the refined solutions at each test period are
shown as red dots in Figure \ref{fig:periodograms}. In addition to the planetary
signals, the period search model also includes one RV zero-point for each
dataset ($\gamma_{HARPS}$ and $\gamma_{PFS}$) and a linear trend (e.g., caused
by very long period companions). The F-ratio at each proposed period is then
computed as 

\begin{eqnarray}
F[P] = \frac{(\chi^2_k-\chi^2_{k+1}[P])/(M_{k+1}-M_{k})}{
\chi^2_{k+1}/(N_{obs}-M_{k+1})}\,, \label{eq:fratio}
\end{eqnarray}

\noindent where $M_{k}$ is the number of free parameters of the solution
including k planets ($5 \times k$ from the previous k planets plus one RV offset
for each dataset, plus one linear trend), and $M_{k+1}$ is the number of
parameters when one additional circular orbit is included ($M_{k+1}$ = $M_{k}$ +
2). We call this procedure \textit{recursive periodogram method} because of its
ability to recursively refine the best Keplerian solution at the period search
level. For numerical efficiency purposes, the Keplerian model and the partial
derivatives of the observables we use are based on the recipes given in
\citet{wright:2009}.

The F-ratio of the preferred solution is then used to estimate its FAP analytically using the general methods described in
\citet{cumming:2004}. In the same study, it was shown that such analytic FAP estimates tend to be over-optimistic because some implicit hypotheses of the method are typically not satisfied by the data (e.g., perfectly known uncertainties, Gaussian posterior distributions, strongly uneven sampling, etc.). This statement was made more precise in \citet{anglada_tuomi:2012}, where it was shown that FAP estimates based on the
analysis of residual Doppler measurements suffer from strong biases that can
work either way (increase false-positive rates, reduce sensitivity to the
detection of small signals). Therefore, empirical FAP assessments are always
mandatory when periodogram methods are used to assess the significance of a new
Doppler candidate. To save unnecessary computations, we will only compute
empirical FAPs if the analytic FAP is already found to be lower than 2\%. Such
empirical FAP computation consists on generating a number of synthetic datasets
(or trials) and quantify how many times a false alarm is induced by an
unfortunate arrangement of the noise. Each synthetic data set is obtained by 1)
randomly permuting the residuals to the k-planet fit over the same observing
epochs (keeping membership to each instrument), and 2) adding back the signal of
the k-planet model. We then apply the recursive periodogram method to the
synthetic set and save the highest synthetic F-ratio in a file. Let us note
that, compared to the more classic Lomb-Scarge periodogram of the residuals,
each recursive periodogram can be computationally very expensive (for GJ 221, 
it took 5 minutes to compute a recursive periodogram to search for a
second signal). While this is reasonably quick on the real data, it
becomes computationally very intensive if 10$^3$--10$^4$ trials are necessary to
confirm the significance of a detection. Fortunately, each trial is independent
of the others making this task very easy to parallelize. We arbitrarily choose
our detection threshold at a FAP of 1\%. This threshold keeps the required
amount of trials down to a reasonable number while keeping the sample of RV
candidates relatively clean of false positives. To save unnecessary
computations, we first run $10^3$ trials. If no FAPs are found, we assume that
the FAP is significantly lower than 1\% and accept the candidate as significant.
If the empirical FAP estimated from the first 10$^3$ trials is higher than 2\%
we also stop the simulations and reject the candidate. Only if the FAP is within
0.1\% and 2\%, we then run 10$^4$ trials and reevaluate the FAP. If the derived
FAP is within 0.5\% and 1\%, we extend the simulations to 5 10$^4$ trials and
accept the final result as the empirical FAP. For this system in particular,
10$^3$ trials to search for a second planet running on 40 logical processors
were obtained in about 4 hours and 10$^4$ were obtained in $\sim 2$ days. 

\begin{figure}
\includegraphics[width=5.0in,clip]{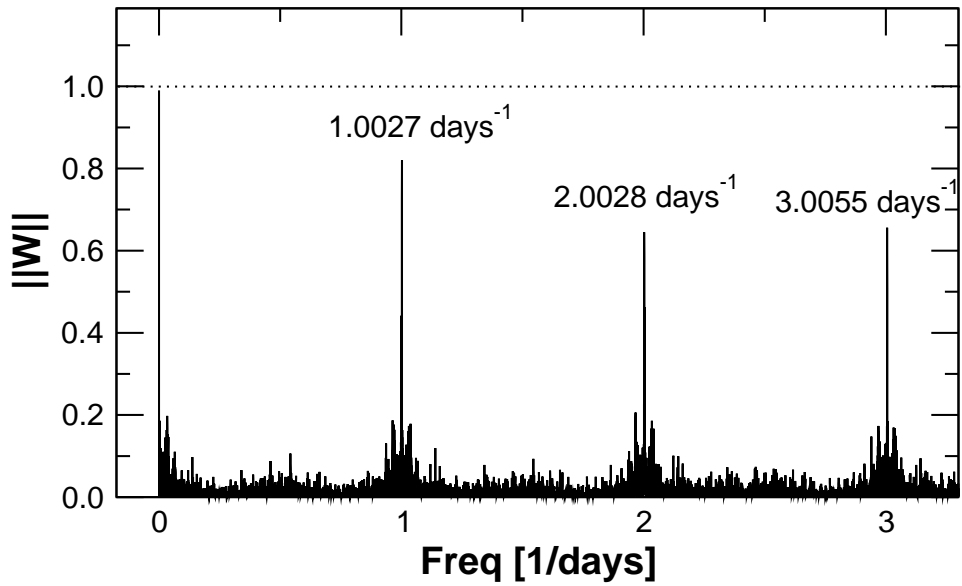}
\caption{Modulus of the window function for the combined HARPS-TERRA + PFS
datasets. Stronger aliases are expected to appear at $\pm$1.0027 and 
$\pm$2.0028 days$^{-1}$ from real signals. Higher order aliases will also
be present at very high frequencies (sub-days period).}
\label{fig:window}
\end{figure}

The sampling cadence can also cause confusion between real periodic signals and
their corresponding aliases. The modulus of the window function \citep{dawson:2010}, is shown in
Fig.~\ref{fig:window}. The window function is strongly peaked at the a few
strong sampling frequencies $f_s$ of the time-series (related to sidereal day).
For a real signal of period $P_k$, very strong aliases will appear at periods
satisfying $1/P_a = \left|1/P_k + f_s\right|$ . The stronger sampling
frequencies in this case are $f_s = \pm 1.0027, \pm 2.0028, \pm 3.0055$
days$^{-1}$ (see Fig.~\ref{fig:window}). Note that the first f$_s$ reaches a
modulus of 0.81, meaning that a perfectly sinusoidal signal will always be
accompanied by two peaks of $\sim$ 80\% of its power. The exact ratio will
depend on interference with additional signals and the 
spectrum of the noise. The top panel of Figure \ref{fig:periodograms} shows the
recursive periodogram search of the combined data sets and the empirical FAP
thresholds for the first candidate. Such periodogram identified a first
candidate with an extremely low analytic FAP ($<$10$^{-10}$, P$\sim$ 125 days). As seen in the top
panel of Fig.~\ref{fig:periodograms}, it is accompanied by its corresponding aliases
around one day corresponding to $1/P_a= 1/125.06 \pm 1.0027$ days $^{-1}$. Since no
false alarms were found in the first 10$^3$ trials the candidate was quickly accepted. Using the stellar parameters
listed in table \ref{tab:starparam}, the best Keplerian fit yields a minimum mass of 0.16 \mjup~ in a slightly eccentric orbit $e=0.17$. The
recursive periodogram search for a second planet (Fig.~\ref{fig:periodograms},
bottom panel), revealed a second dominant signal at 3.8741 days. Very prominent
alias of this signal are also clearly detected at 1.34285 days (f$_s$= -1.0027
days$^{-1}$), 0.79307 days (f$_s$= +1.0027 days$^{-1}$) and 0.5731 days (f$_s$=
-2.0028 days$^{-1}$). The signal is apparent both in the periodogram of the
residuals (black line) and in the recursive periodogram refinement (red dots).
The corresponding Keplerian solution results in a minimum mass of \msini of 6.34
\mearth in a slightly eccentric orbit ($e=0.17$). The empirically estimated FAP
required 10$^4$ trials and was found to be about $\sim$ 0.44\%. Therefore, this
candidate also satisfied our detection criteria and was added to the solution.
The three horizontal lines in both figures represent are the 10\%, 5\%, and
1.0\% FAP levels as derived from the empirical FAP calculations. No candidates
with an analytic FAP lower than 2\% appeared in the recursive periodogram search
for a 2+1-planet solution.

Confidence intervals for the model parameters were computed using the Bayesian
Monte Carlo Markov Chain method(MCMC) explained in \citet{ford:2005} but with a
slightly different choice of free parameters. The model parameters describing
one Keplerian orbit are always referred to some reference epoch $t_0$ (which we
arbitrarily choose on the first date of the HARPS observations) and they are :
the period $P$, semi-amplitude $K$, orbital eccentricity $e$, the initial mean
longitude $\lambda_0$, and the argument of the periastron $\omega$ (angle between 
the periastron of the orbit and the ascending
node). The angle $\lambda_0$ is defined as $\lambda_0 = M_0+\omega$ and its
choice instead of $M_0$ (mean anomaly at $t_0$) is justified as follows. 
When the orbit is close to circular, the position of the periastron 
with respect to the plane of the sky is ill-defined and uncertain. 
As a result, $\omega$ and $M_0$ are strongly degenerate
for small eccentricities. Similarly in the time domain, the precise orbital
phase at which the periastron is crossed (instant $t$ for which $2\pi/P
\left(t-T_0\right) + M_0 = 2\pi\times N$, where $N$ is an integer number) is
also very uncertain. As has been discussed in previous works \citep[e.g.,
][]{gj1214}, the sum $\lambda_0$ contains all the relevant information about the
orbital phase when the orbit is close to circular and is much better behaved
(non-degenerate) than the two other angles separately. In a technical sense, the
convergence of the Markov Chains is also greatly improved thanks to the
elimination of the degeneracy between $M_0$ and $\omega$ at any $e$. Let us note
that the best fit values for M$_0$ can be trivially recovered using of $\omega$ and
$\lambda_0$ from table \ref{tab:solution}.

\section{Activity : CaII H+K}
\label{sec:activity}

We limit our analysis of the activity to searching for signals in the S-index
with a period coincident (or close to) any of the RV candidates. Some other
activity indicators are provided by the HARPS-DRS (bisector span,
full-width-at-half-maximum of the cross correlation function), but these indices
could only be examined on 51 HARPS observations that produced meaningful CCF RV
values. In any case, a quick-look at their periodograms indicated that none of
them showed a periodicity with an analytic FAP lower than 40 \%.

The HARPS-TERRA software computes the S-index in the Mount Wilson system using
the prescription given by \citet{lovis:2011} on the blazed corrected
1-dimensional spectrum as provided by the HARPS-Data Reduction Software. In the
case of PFS, instrumental S-indices were derived from the blaze-corrected
spectra following the definition given by \citet{baliunas:1995}. We only used
the CaII H line, because the K line is in a spectral order with very low typical
fluxes, which introduces undesirable noise. To convert these values to the
Mount Wilson standard system, a simple linear calibration between the PFS and
Mt. Wilson systems was made. The main problem of calibrating the S scale is that
a number of standard stars observed with both instruments are needed. To address
this, we turned to our previously measured values \citep{arriagada:2011} for some
of these stars, performed a linear least-squares fit and applied the relation
to the present measurements.

We computed the least-squares periodogram of the combined series and found a
significant periodic signal at 1.84 days (and a corresponding daily alias at
2.16 days) plus strong power at long periods, so we included the adjustment of
the trend at the period search level (as we did for the RVs). It is likely that
the long period trend is due to the stellar magnetic cycle of the star (several
years?). However, the origin of the 1.84 day signal is less clear. The rotation
period of the star could be responsible for such a signal but this would make GJ
221 a very fast rotator and likely to suffer from strong RV jitter \citep[see
e.g., TW Hya in ][]{setiawan:2008}. When the 1.84 day signal is removed, a
significant periodicity remains at 15 days, again with no RV counterpart. This
would be within the range of the expected rotational period of a main sequence
K7 star \citep{engle:2011}. In any case, none of the periods detected in the
S-index time series appear to have a counterpart in the RVs. Activity induced
signals acting at different time-scales (1.86 days, 15 days and several years)
are a likely component of the excess noise found in the RVs. In conclusion,
although periodic signals are detected in the S-index, neither candidate has a
period compatible with them, so the Keplerian origin remains the most likely
explanation to the signals in the data.

%In conclusion, although periodic signals are detected in the S-index, none of the proposed two candidates have a period compatible with them so the Keplerian origin remains as the most likely explanation to the signals in the data.

\section{Conclusions} %or discussion? or summary?

We have analyzed combined precision RV measurements of the K/M dwarf GJ 221
obtained using Magellan/PFS observations and HARPS public spectra. The
HARPS-TERRA measurements combined with 17 additional PFS observations reveal the
presence of two low-mass companions. 

\begin{figure}
%\plotone{stable1.ps}
\includegraphics[width=5.0in]{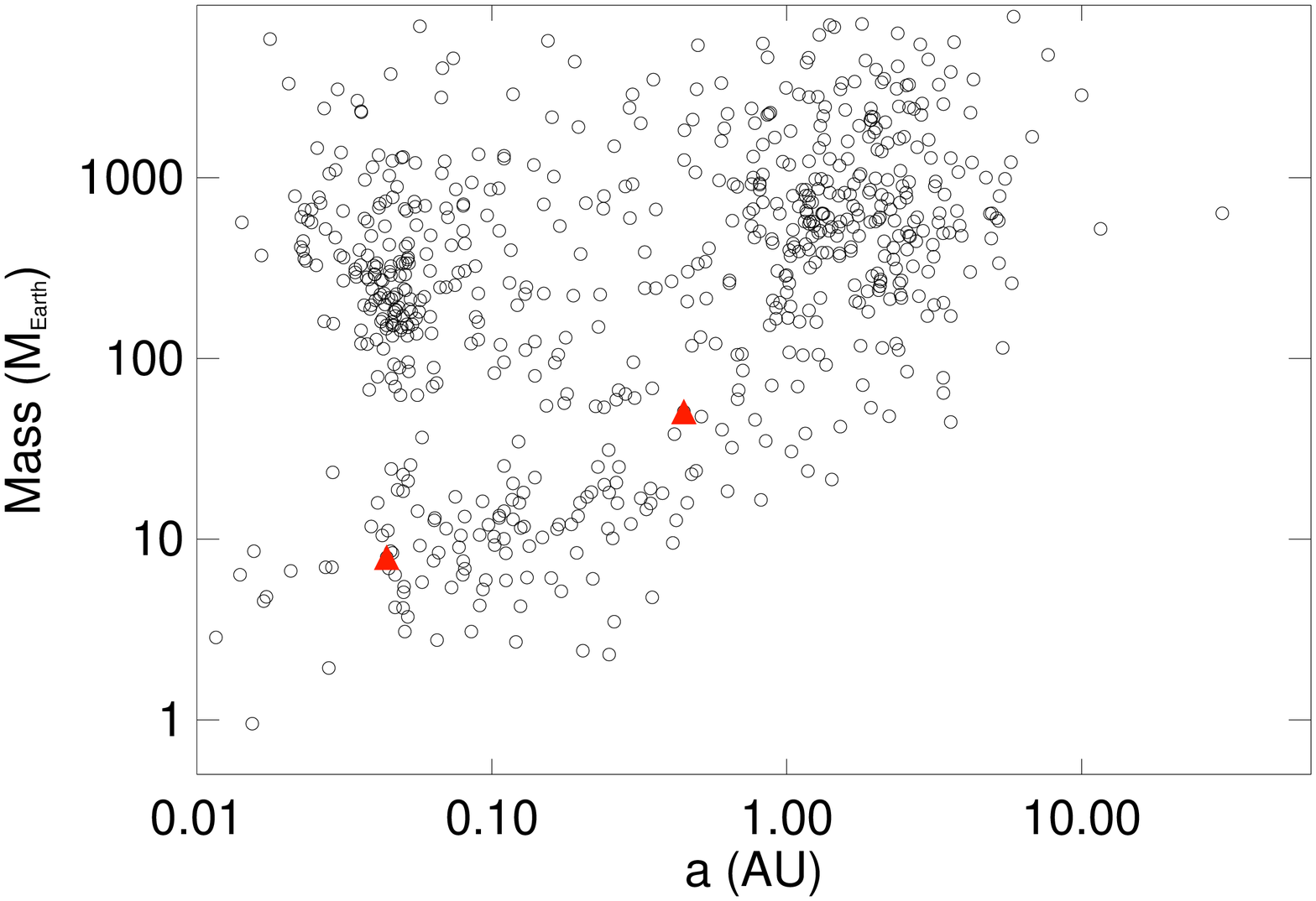}
\includegraphics[width=5.0in]{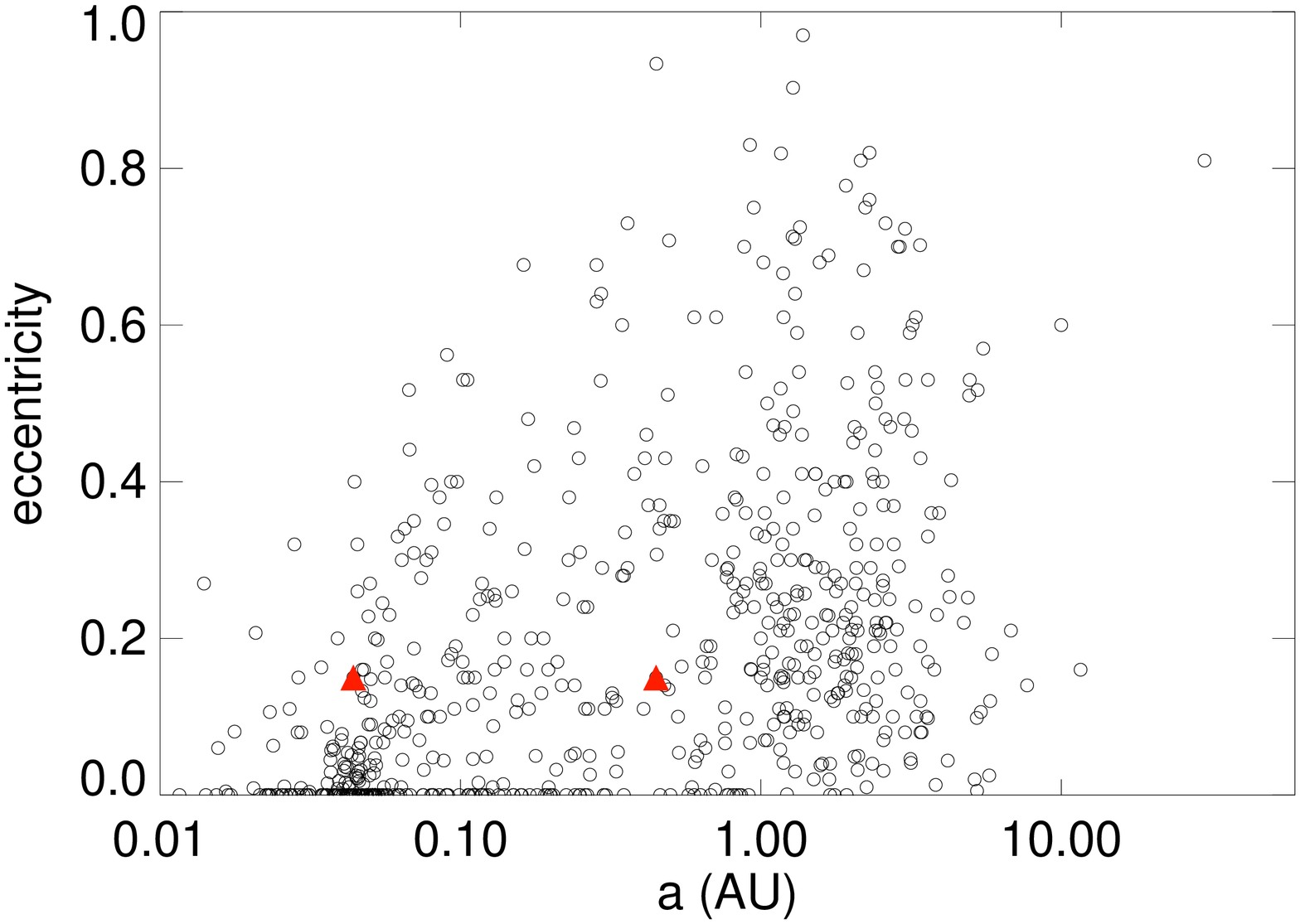}
\caption{Plot of orbital elements of all known exoplanets (open circles) and the new discovered planets presented in this paper (red filled triangles).}
\label{orbital_elements}
\end{figure}

Figure \ref{orbital_elements} shows our planets in the semimajor axis-mass and
semimajor axis-eccentricity parameter spaces of all known extrasolar planets.
All of our new detected planets lie well within the parameter space envelope. 
It is important to note that the longer period planet (GJ221b) lies in
the middle of the so-called giant-planet ``desert'' predicted by population synthesis models
\citep{ida:2004, ida:2010, mordasini:2012}, which is in accordance with other observational evidence \citep{cumming:2008, howard:2012}. %Other studies have found discrepant results regarding this desert, \citet{howard:2010} finds a paucity of Neptunes at $P<20$ days, while \citet{howard:2012} sees no evidence of this paucity in Kepler data. 
Given the low metallicity content of GJ 221, the planetary mass of  GJ221c is well in agreement with what observations show in this mass regime \citep{sousa:2008}, as well as what it is predicted by models \citep{mordasini:2012} in that lower-mass planets do not preferentially form around metal-rich stars. 

Quantifying planet occurrence for M dwarfs provides major constraints for further tunings of these models as well as support for post disk effects such as secular planet migration or planet-planet interactions.

Using the relation given by \citet{charbonneau:2007}, the
companion with shorter period (GJ 221b) has a significant probability (8\%) of
transiting in front of the star and deserves further photometric follow-up.
Finally, the star is slightly metal depleted compared to the Sun
([M/H]$\sim$-0.26) giving further indication that low metallicity \citep[e.g., ][]{anglada:2012b}  does not inhibit
the formation of planets in the super-Earth/Neptune mass regime.

\acknowledgements

DM and PA have been supported by the Basal Center for Astrophysics and 
Associated Technologies (CATA) PFB-06, and by the Ministry for
the Economy, Development, and TourismÕs Programa Iniciativa Cient\'ifica Milenio 
through grant P07-021-F, awarded to The Milky Way Millennium Nucleus. GAE is supported by the German Federal Ministry of
Education and Research under 05A11MG3. We are grateful for the advice,
support and useful discussions with Ansgar Reiners (IAG),
Mathias Zechmeister (AIG) and Hugh Jones (HU). We thank Sandy Keiser
for setting up and managing the computing resources available at
DTM/CIW. We are grateful for the support that Frank H. Pearl
provided at a critical time in the development of the Planet Finding
Spectrograph. This work is based on data obtained from the ESO Science
Archive Facility under request number GANGLFGGCE175695. This research
has made extensive use of the SIMBAD database, operated at CDS,
Strasbourg, France; and the NASA Astrophysics Data System.

\clearpage

\clearpage

\clearpage

\clearpage

 \clearpage
\begin{figure}
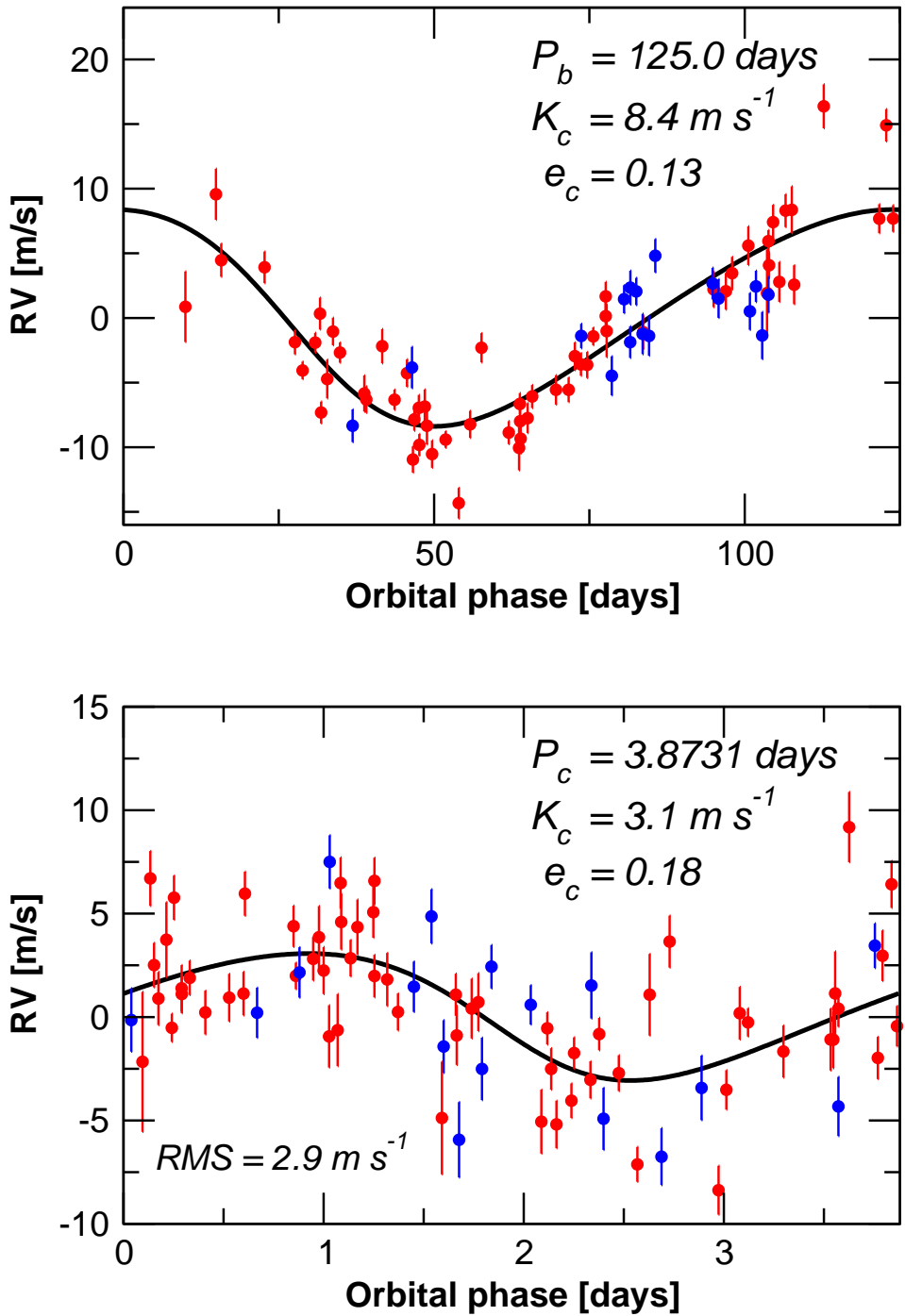

%\plotone{stable1.ps}
\includegraphics[width=5.0in]{GJ221_b_phased.eps}
\vskip1.0cm
\includegraphics[width=5.0in]{GJ221_c_phased.eps}
\caption{Keplerian solution for GJ221. Top panel: phased Keplerian fit for the 125 day component. Bottom panel: phased Keplerian fit for the 3.87 day component. Red symbols correspond to HARPS velocities, while blue circles correspond to PFS velocities.}
\label{keplerian}
\end{figure}

\clearpage

\clearpage
\begin{figure}
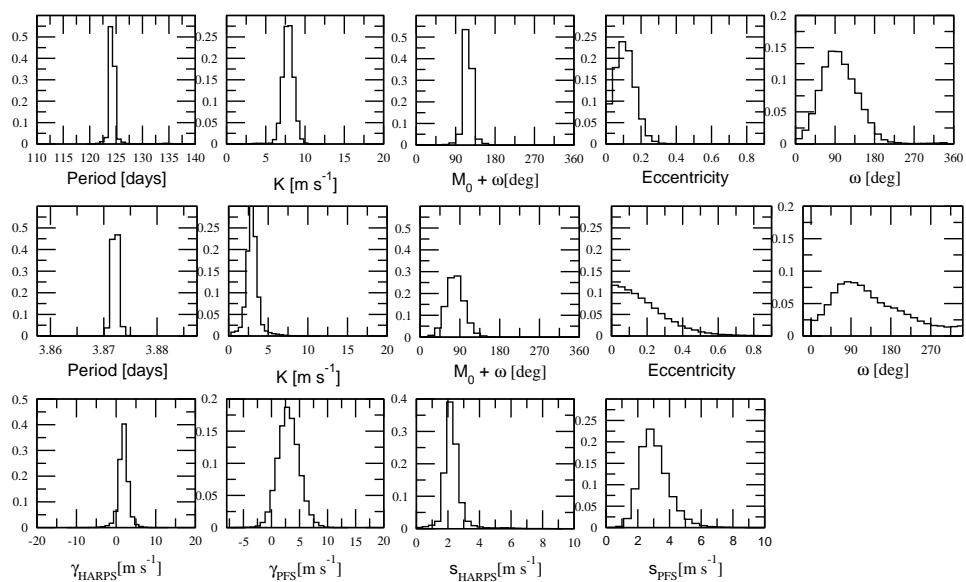

\includegraphics[width=5.0in,clip]{Hist_b.eps}
\includegraphics[width=5.0in,clip]{Hist_c.eps}
\includegraphics[width=4.0in,clip]{Hist_Extra.eps}
\caption{Marginalized posterior distributions for all the parameters in the
model. First two rows correspond to the orbital elements of GJ 221b and GJ 221c.
Last row contains the data set-dependent parameters such as the zero-point
offsets (two leftmost panels) and both jitter parameters (two rightmost panels).}
\label{fig:mcmc}
\end{figure}

\clearpage

\begin{deluxetable}{rrr}
\tablecaption{HARPS-TERRA Velocities for GJ 221}
\tablewidth{0pt}
\tablehead{
JD & RV & error \\
(-2400000)   &  (m s$^{-1}$) & (m s$^{-1}$) 
}
\startdata
53288.88 & -2.72 & 1.02 \\
53311.82 & -3.88 & 0.89 \\
53314.86 & -2.44 & 1.15 \\
53373.68 & 12.96 & 1.03 \\
53403.65 & 0.40 & 0.68 \\
53405.66 & -1.94 & 0.76 \\
53409.66 & -3.06 & 0.78 \\
53438.60 & -1.99 & 0.83 \\
53440.60 & -6.40 & 0.90 \\
53728.74 & 7.27 & 0.98 \\
53781.66 & -8.37 & 0.84 \\
53782.63 & -2.84 & 1.50 \\
53783.57 & 3.73 & 1.06 \\
53788.61 & -3.68 & 1.40 \\
53833.49 & 2.54 & 1.28 \\
54122.65 & 13.55 & 1.26 \\
54140.61 & 8.45 & 1.28 \\
54166.53 & 0.86 & 1.33 \\
54168.52 & -6.85 & 0.78 \\
54170.52 & -0.87 & 1.07 \\
54172.51 & -10.89 & 0.84 \\
54174.55 & -6.71 & 1.06 \\
54194.51 & -1.09 & 1.14 \\
54196.52 & -6.25 & 0.95 \\
54197.54 & 0.10 & 1.08 \\
54198.52 & 0.74 & 0.88 \\
54199.52 & -4.30 & 1.00 \\
54200.50 & -1.74 & 0.69 \\
54202.51 & 4.23 & 1.23 \\
54202.51 & 5.75 & 1.12 \\
54225.47 & 10.03 & 1.50 \\
54228.46 & 4.75 & 3.38 \\
54229.46 & 11.76 & 1.34 \\
54230.46 & 2.33 & 1.54 \\
54231.45 & 7.79 & 1.27 \\
54232.46 & 11.56 & 1.82 \\
54347.86 & 3.62 & 1.26 \\
54384.88 & 3.30 & 2.72 \\
54389.79 & 7.69 & 1.98 \\
54421.72 & -6.73 & 0.87 \\
54423.76 & -6.86 & 1.45 \\
54426.76 & -5.27 & 0.64 \\
54428.87 & -15.48 & 1.18 \\
54430.71 & -4.08 & 1.05 \\
54438.58 & -5.97 & 1.74 \\
54452.69 & 0.00 & 2.03 \\
54478.72 & 3.90 & 0.86 \\
54487.62 & 17.60 & 1.70 \\
54522.64 & 5.72 & 1.22 \\
54531.55 & 4.26 & 1.24 \\
54557.54 & -0.39 & 1.13 \\
54719.92 & 2.81 & 1.38 \\
54721.91 & 3.58 & 1.42 \\
54732.89 & 6.25 & 1.50 \\
54813.80 & -4.69 & 1.05 \\
54871.59 & 10.19 & 1.12 \\
54902.57 & 0.59 & 0.93 \\
54921.54 & -9.80 & 1.02 \\
54922.49 & -3.52 & 1.00 \\
55298.50 & -4.15 & 1.33 \\
55438.93 & -11.90 & 1.15 \\
\enddata
\label{tab:TERRAvels}
\end{deluxetable}

\clearpage

\begin{deluxetable}{rrr}
\tablecaption{PFS Velocities for GJ 221}
\tablewidth{0pt}
\tablehead{
JD & RV & error \\
(-2400000)   &  (m s$^{-1}$) & (m s$^{-1}$) 
}
\startdata

55198.67 & -1.87 & 0.93 \\
55581.60 & 3.77 & 1.32 \\
55671.48 & -6.34 & 1.61 \\
55786.94 & -6.15 & 1.26 \\
55844.88 & 4.53 & 1.27 \\
55845.83 & -1.00 & 1.59 \\
55850.88 & 0.52 & 1.45 \\
55851.85 & 5.68 & 1.27 \\
55852.86 & -0.30 & 1.88 \\
55853.87 & -0.61 & 1.46 \\
55953.67 & -3.59 & 1.54 \\
55955.63 & 2.31 & 1.12 \\
55956.63 & 0.81 & 1.24 \\
55957.59 & 1.57 & 1.05 \\
55958.64 & -4.20 & 1.58 \\
55959.66 & 1.13 & 1.57 \\
55960.66 & 7.96 & 1.31 \\
56282.71 & -0.19 & 1.19 \\
56284.69 & -5.89 & 1.13 \\
56291.73 & -9.73 & 1.14 \\
56354.55 & 7.01 & 1.34 \\
56356.56 & 0.00 & 1.12 \\
56358.55 & 2.47 & 1.49 \\
\enddata
\label{tab:PFSvels}
\end{deluxetable}

\clearpage

\begin{deluxetable}{lcrc}
\tablecaption{Stellar Properties of GJ 221 (HIP 27803)}
\tablewidth{0pt}
\tablehead{
\colhead{Par.} & 
\colhead{Units} & 
\colhead{GJ 221 (HIP 27803)} &
\colhead{Ref.}
} 
\startdata
  R.A.        &  [HHMMSS.SSS]   & 05 53 00.284       & (1)  \\
  Dec.        &  [ddmmss.ss]    &-05 59 41.43        & (1)  \\
  $\pi $      & [mas]           & 49.23 $\pm$1.65    & (1)  \\
  $\mu_{R.A.}$&[mas yr$^{-1}$]  &  -1.08$\pm$1.55    & (1)  \\
  $\mu_{Dec}$ & [mas yr$^{-1}$] &-346.17$\pm$1.32    & (1)  \\
  $B$         & [mag]           & 11.04$\pm$0.01     & (2)  \\
  $V$         & [mag]           & 9.693$\pm$0.01     & (2)  \\
  $K$         & [mag]           & 6.305$\pm$0.005    & (3)  \\
%  $M_V$       & [mag]           &  8.15              & (*)  \\
  Hel. RV         & [km s$^{-1}$] & 40.1             & (*)\\
  \\
  \multicolumn{3}{l}{Derived quantites}\\
  \hline
  & & \\
  UVW$_{\rm{LSR}}$ & [km s$^{-1}$] & (-16,-40,-23) & (*) \\
  Age              & [Gyr]         & 1--5          & (*) \\
  & & \\
  T$_{eff}$ & [K]              &   4040 $\pm$ 50    & (*)    \\
  log g     & g in [cm s$^{-1}]$ & 4.5  $\pm$ 0.1   & (4,*)  \\
  $[Fe/H]$  & [dex]            &  -0.07 $\pm$ 0.10  & (*)    \\
  $v_{rot}\sin{i}$
            & [km s$^{-1}$]    &   1.8  $\pm$ 0.1   &  (*)\\
  & & \\
  M$_*$     & [\msun]          & 0.637 $\pm$ 0.032 & (5,*)\\
  L$_*$     & [L$_\sun$]       & 0.095 $\pm$ 0.003 & (5,*)\\
% $\theta$ &[mas]  &0.312(0.019)\\
% M$_*$ & [\msun]  & 0.767         & (3) \\
\enddata
\label{tab:starparam}
\tablecomments{
(1) \citep{hipparcos:2007},
(2)\citep{koen:2010}, 
(3) \citep{twomass}, 
(4) \citep{takagi:2011},
(5) Evolutionary models from \citep{baraffe:1998},
(*) This work (see text)
}
\end{deluxetable}

\clearpage

\begin{deluxetable}{rll}
\tablecaption{Orbital Parameters corresponding to the model that maximizes
the a posteriori probability. Uncertainties correspond to the standard 
deviation of the marginalized MCMC samplings.}
\tablewidth{0pt}
\tablehead{
\colhead{Planet}  & \colhead{GJ 221b} & \colhead{GJ 221c} } 
\startdata 
P [days]         & 125.06 (1.1)       & 3.8731(6.6 $10^{-4}$)\\
K[\ms]	         & 8.35 (0.76)        & 3.15  (0.87)         \\
$M_0$+$\omega$ [deg]& 131 (17.5)        & 95.9 (20.0)         \\ % con e_c=0
$e$              & 0.13	($<0.24^{*}$) & 0.18 ($<0.48^{*}$)   \\
$\omega$ [deg]	 & 118 (44)        & 94 (unc)            \\
$\gamma_{HARPS}$ &       2.9 (1.3)    &                      \\
$\gamma_{PFS}$   &       4.1 (1.9)    &                      \\
\\
$\left<s_{HARPS}\right>$  
                 &    2.6 (0.62)      &       \\
$\left<s_{PFS}\right>$    
                 &    3.5 (1.9)       &       \\
Derived quantities		 \\
\hline\hline\\
M$_*$[\mjup]     &  0.13              & 0.0196      \\
a[AU]            &  0.448             & 0.044       \\
Detection FAP	 & $<<0.1\%$          & 0.44\%      \\
RMS [\ms]        &    2.86 &                        \\
\enddata
\tablenotetext{*}{Eccentricity poorly constrained, 95\% c.l. Upper limit given instead.}
\label{tab:solution}
\end{deluxetable}

\clearpage

\end{document}